\documentclass{article}
\usepackage{graphicx}
\usepackage{geometry}
\usepackage[font=small,labelfont=bf]{caption}
\usepackage{float}
\usepackage[nottoc,numbib]{tocbibind}
\usepackage[sort&compress,numbers]{natbib}
\usepackage{siunitx}
\usepackage{booktabs}

\begin{document}
\begin{titlepage}
	\centering{

	{\Large\bfseries Near-perfect absorber as a subwavelength thickness spatial frequency filter for optical image processing\par}
	\vspace{0.5cm}
		
		{\large\itshape Lukas Wesemann $^1$, Evgeniy Panchenko $^1$, Kalpana Singh $^1$,\par Enrico Della Gaspera $^2$, Daniel E. G\'omez $^2$, Timothy J. Davis $^1$ \par and Ann Roberts $^1$ \par}
		\vspace{1cm}
		{\small $^1$ School of Physics, The University of Melbourne, 3010, Australia \par $^2$ School of Science, RMIT University, GPO Box 2476, Melbourne VIC 3001 Australia\par}
			\vspace{0.5cm}
		{\scshape\large $15{^\mathrm{th}}$ February 2019\par}
}

\end{titlepage}

\section{Abstract}
Spatial frequency filtering is a fundamental enabler of information processing methods in biological and technical imaging. Most filtering methods, however, require either bulky and expensive optical equipment or some degree of computational processing. Here we experimentally demonstrate on-chip, all-optical spatial frequency filtering using a thin-film perfect absorber structure. We give examples of edge enhancement in an amplitude image as well as conversion of a phase gradient in a wave field into an intensity modulation.

\section{Introduction}
Spatial frequency filtering underpins many  widely used imaging methods ranging from visualizing phase variations in transparent objects such as live biological cells \cite{stephens2003light} to approaches for image enhancements used for object and face recognition \cite{HJELMAS2001236}.
Most common optical approaches to spatial frequency filtering, however, involve relatively bulky optical systems \cite{zernike1942phase, murphy2001differential} while other popular image processing techniques rely on computation \cite{HJELMAS2001236}.
While the use of conventional optical components places limits on the potential for miniaturization, the conversion from optical to electrical signals as well as computation time for electronic approaches places limits on applications requiring high-throughput and rapid processing \cite{chen2014data, pham2000current, solli2012fluctuations}. Furthermore computational approaches have greater power requirements.\\ \\
Nanophotonic, all-optical image processing methods on the other hand offer ultra-compact, real time and low power consumption alternatives for use in emerging integrated analogue optical computing and image processing devices \cite{solli2015analog,silva2014performing}.
\begin{figure}[h]
\centering
\includegraphics[width=0.7\linewidth]{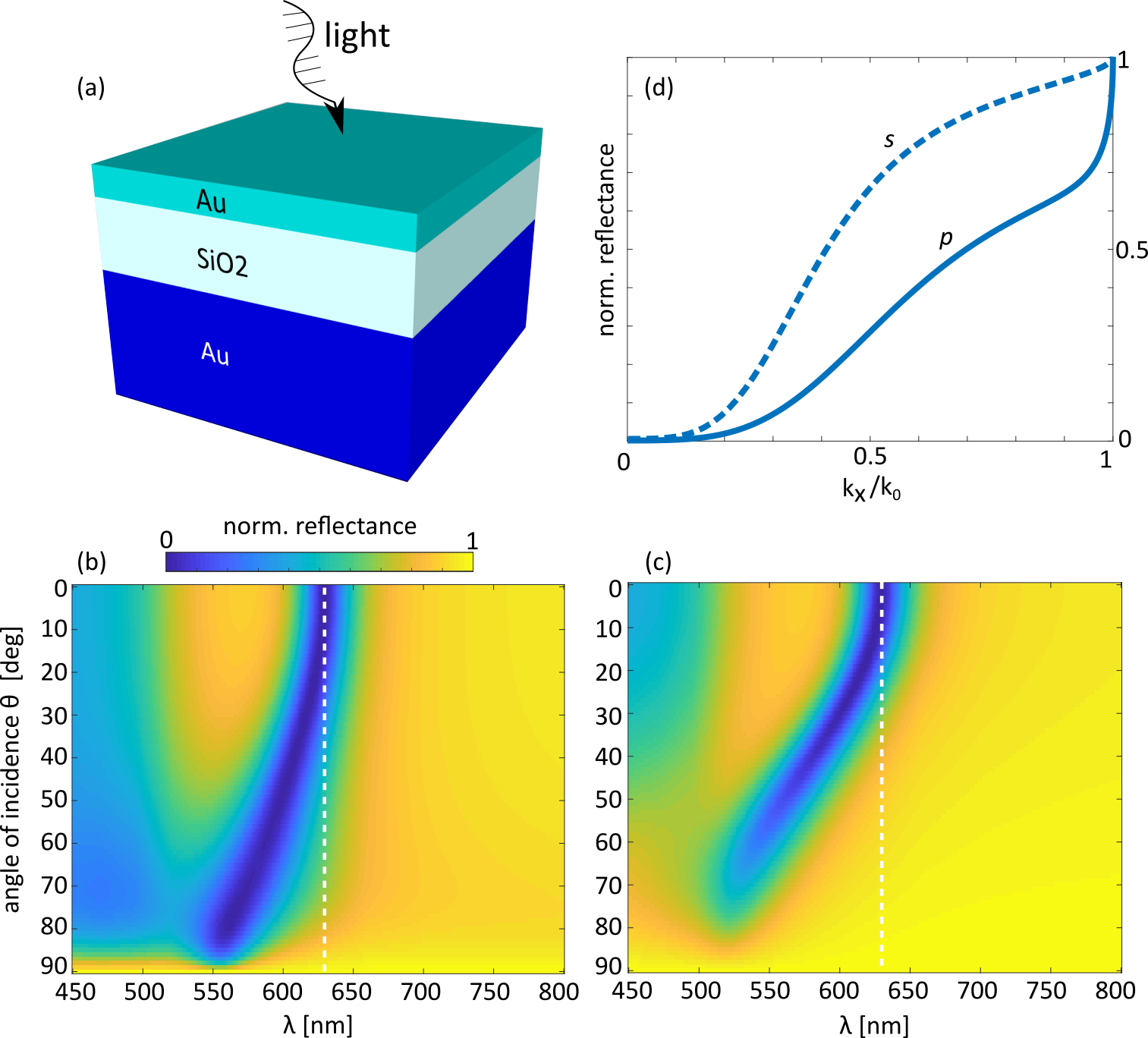}
\caption{Salisbury screen as a highpass spatial frequency filter. (a) Thin film configuration, (b),(c) calculated reflectance spectrum as a function of angle of incidence for \textit{p-} and \textit{s}-polarization, (d) calculated reflectance for \textit{p}- (solid) and \textit{s}-polarization (dashed) evaluated at the absorption wavelength $\lambda=631$nm.}
\label{fig:fig1}
\end{figure}
There has been considerable recent interest in the use of metasurfaces or subwavelength gratings for optical information processing \cite{silva2014performing,roberts2018, Hwang_Optical_2016, Davis_A_2009, Guo:18}. Attention has focussed on optical computing of spatial differentiation although experimental demonstrations have been limited and restricted to operations in one dimension \cite{bykov18,Dong_Grating_2018}. While nanophotonic devices usually require complex fabrication schemes, thin-film approaches have recently gained attention as solutions for all-optical image processing \cite{Wu_Multilayered_2017,golovastikov2015spatial,bykov2014optical} although suggested designs are usually of the size of several wavelengths. Experimental demonstrations to date are still relatively bulky and have additionally been confined to performing operations in one dimension \cite{Zhu_Plasmonic_2017}.
Inspired by the concept of the Salisbury screen developed by Winfield W. Salisbury \cite{2599944_Absorbent_1952} we here report a simple two layer thin-film device that performs all-optical 2D spatial frequency filtering of an incoming wave field, with the relevant thickness of the device being less than $200$nm. This spatial frequency filtering capability of thin film absorbers suggests an avenue for the development of ultra-compact, all-optical information processing devices. Due to their relatively small footprint and simple fabrication process, thin film absorber devices have the potential to be part of next generation image processing components in handheld medical diagnostic and other portable sensing systems \cite{Breslauer_Mobile_2009}. Here we demonstrate experimentally the edge enhancement of an amplitude image and conversion of a phase gradient into intensity modulation using a thin-film absorber.
\section{Spatial frequency filtering capability of thin film absorbers}
\begin{figure}[H]
\centering
\includegraphics[width=0.6\linewidth]{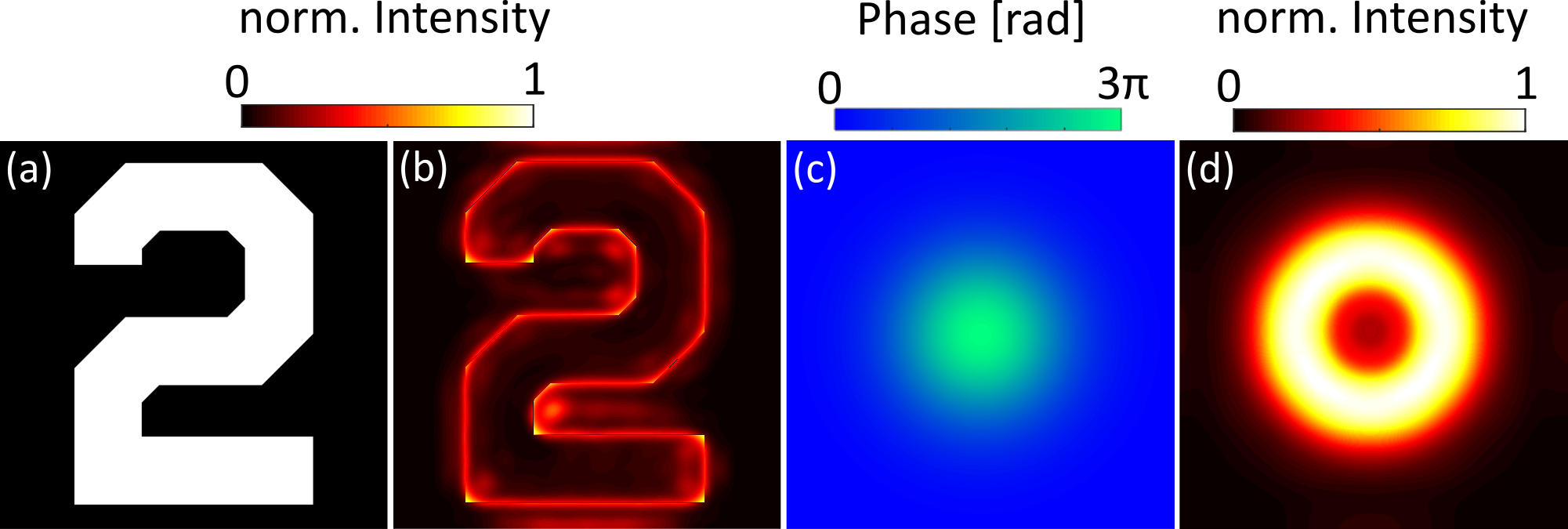}
\caption{Numerical demonstration of all optical image processing by the thin film absorber structure at $\lambda=631$nm using unpolarized light. Edge enhancement in an amplitude image with (a) the original and (b) the intensity of the reflected image. Visualization of a phase-only gradient with (c) the phase modulation and (d) the intensity of the reflected image. The image is $6\times6\mu$m in size.}
\label{fig:impronum}
\end{figure}

A Salisbury screen consists of a thin, absorbing, but semitransparent, layer of material separated from a reflective substrate by a dielectric spacer layer. For a dielectric layer with a thickness of approximately a quarter wavelength multiple reflections lead to 'perfect' absorption at a specific wavelength. The condition for absorption is, however, also sensitive to angle of incidence so that the reflectance depends on the transverse component of the spatial frequency of an incident plane wave. As an example consider a device consisting of a gold film with thickness $d_{1}=31$nm and a silicon dioxide spacer layer of thickness $d_{2}=155$nm on a semi-infinite gold substrate as illustrated in Figure \ref{fig:fig1}(a). Thicknesses were chosen to minimize reflectance at normal incidence at a wavelength of $\lambda=631$nm, although grazing angles are strongly reflected. The optical properties of bulk gold \cite{johnchristy72} and fused silica \cite{malitson1965} were taken from the literature. The reflectance spectrum from the device as a function of angle of incidence is shown for both \textit{p}- and \textit{s}- polarized light in Figures \ref{fig:fig1}(b) and (c) respectively. Negligible reflection can be seen at normal incidence, but the reflectance increases with angle of incidence (\textit{c.f.} Fig. \ref{fig:fig1}(d)). This structure, therefore, suppresses plane waves with low spatial frequencies while reflecting high spatial frequencies at the design wavelength demonstrating the capacity for spatial frequency filtering. This property holds for both \textit{p-} and \textit{s-} polarization with only small differences between the two in sensitivity of the reflectance to $k_x$. Figure \ref{fig:impronum} shows the results of numerical simulation of image processing by reflection from a Salisbury screen of both an amplitude and a pure phase object. In the case of plane wave illumination of the object of interest, the resulting transmitted wavefield incident onto the screen can be decomposed into plane wave \textit{p}- and \textit{s}-polarization components with $\tilde{E}_{0p}(k_x,k_y)$ and $\tilde{E}_{0s}(k_x,k_y)$ denoting their Fourier transforms respectively. The amplitude reflectances for \textit{p}- and \textit{s}-polarization are given by $r_p(k_x,k_y)$ and $r_s(k_x,k_y)$ and these play the role of optical transfer functions. The \textit{p}- and \textit{s}-components of the angular spectrum of the reflected wavefield are, therefore, given by:

\begin{eqnarray}
\tilde{E}_{\mathrm{ref},p}(k_x,k_y)=r_p\tilde{E}_{0p}(k_x,k_y) \\
\tilde{E}_{\mathrm{ref},s}(k_x,k_y)=r_s\tilde{E}_{0s}(k_x,k_y)
\label{eq:refname1}
\end{eqnarray}

Inverse Fourier transforming gives the \textit{p}- and \textit{s}-components of the reflected fields $E_{\mathrm{ref},p}(x,y)$ and $E_{\mathrm{ref},s}(x,y)$ from which the intensity of the output field can be calculated. Here we assume unpolarized illumination and ignore any polarization effects introduced by the object.
The angular sensitivity of the plane wave reflectance from the screen and the one-to-one correspondence between spatial frequencies $k_{x},k_{y}$ and its propagation direction enhances regions in the reflected (output) image associated with high spatial frequencies, e.g. sharp edges in an amplitude object or regions with strong phase gradients, while regions dominated by low spatial frequencies, i.e. homogeneous areas, will appear darker. 
Figures \ref{fig:impronum}(a,b) show an amplitude object and the calculated intensity reflected from the screen at $\lambda=631$nm. The image incident on the screen is given dimensions of $6\times6\mu$m. It is evident that edges in the reflected image (Fig. \ref{fig:impronum}(b)) appear bright compared to regions dominated by low spatial frequencies thus demonstrating the edge enhancement capabilities. Figure \ref{fig:impronum}(c,d) further demonstrate the conversion of phase gradients of an incident pure phase object to variation in intensity in the processed image. The input has a total phase excursion of $\Delta \varphi_{\mathrm{max}}=3\pi$ radians in (c) with the resulting image in (d) exhibiting intensity variations associated with regions of higher phase gradient. These results demonstrate the potential of the structure to suppress low spatial frequencies in the reflected field enabling edge enhancement and conversion of phase gradients into intensity variations.

\section{Experimental Results}
\begin{figure}[H]
\centering
\includegraphics[width=0.7\linewidth]{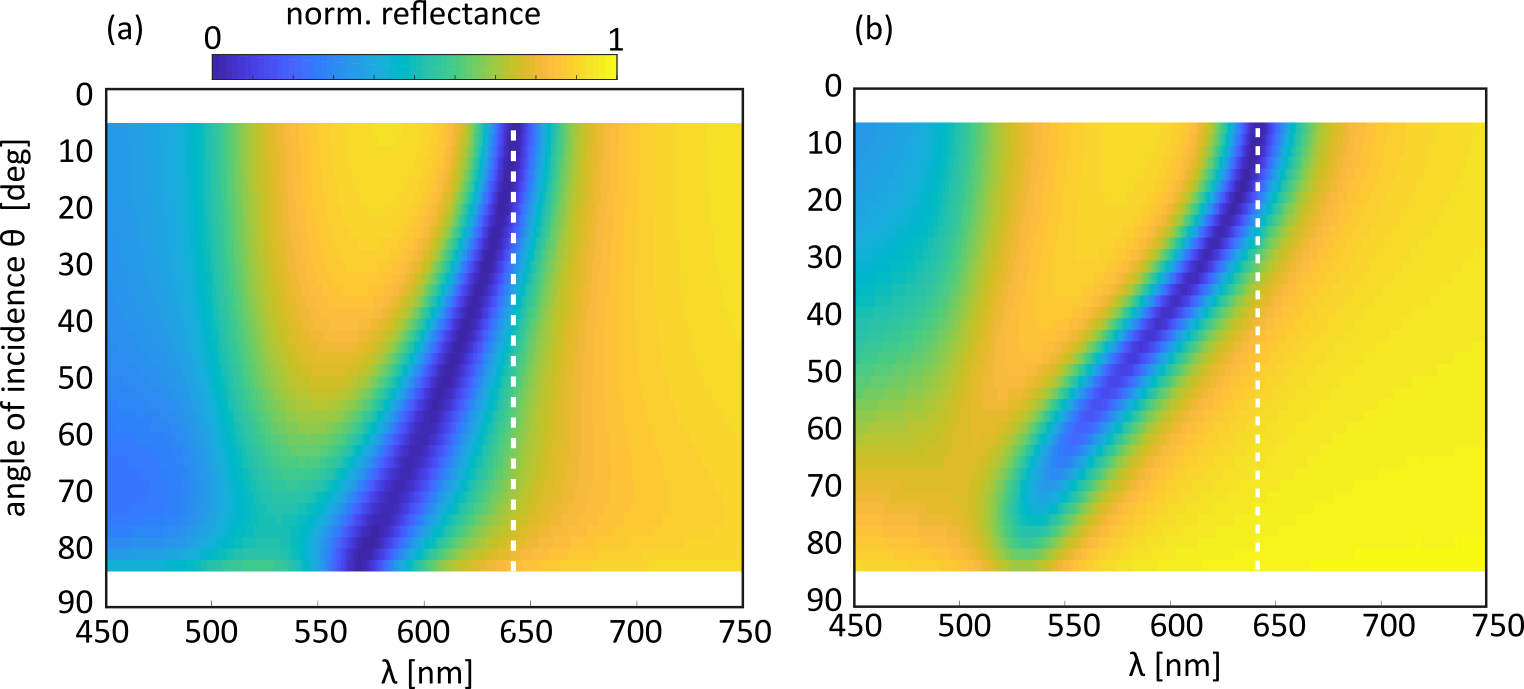}
\caption{Measured reflectance from the device under consideration as a function of wavelength $\lambda$ and angle of incidence for (a) \textit{p-} and (b) \textit{s-}polarization.}
\label{fig:reflectance}
\end{figure}

\begin{figure}[h]
\centering
\includegraphics[width=0.6\linewidth]{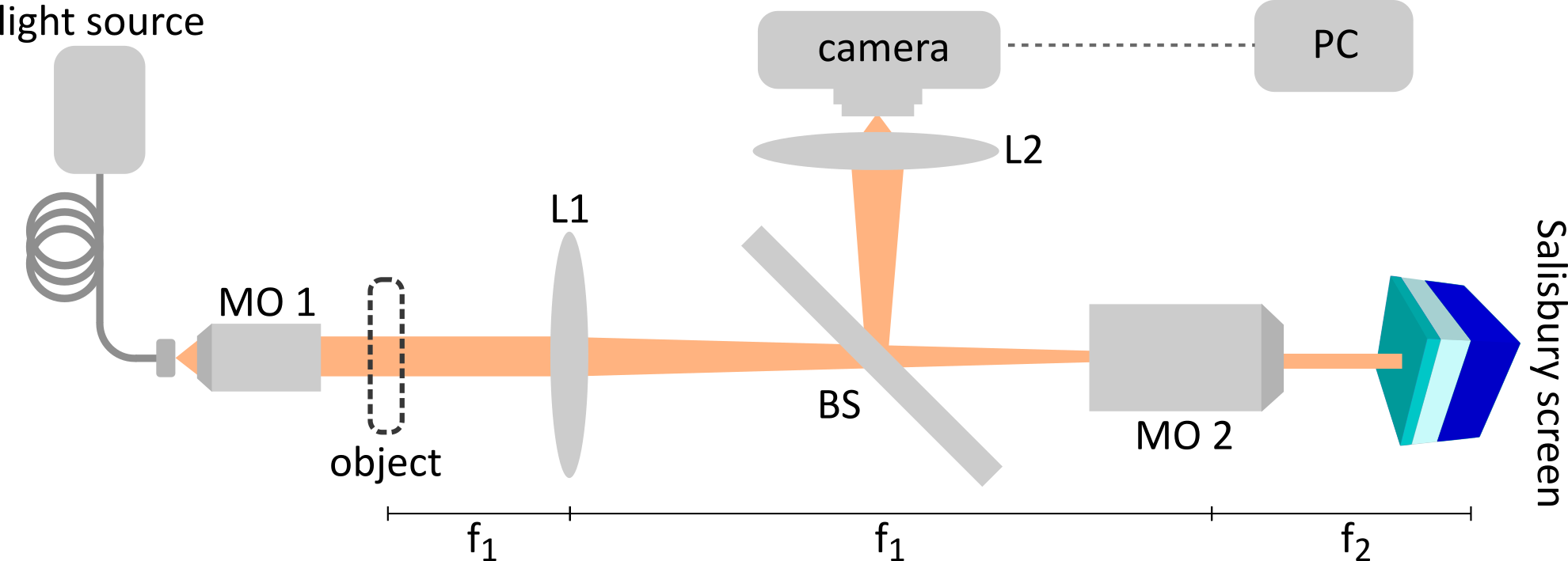}
\caption{K\"ohler-illumination setup used for demonstration of edge enhancement and phase gradient detection.}
\label{fig:setup}
\end{figure}

\begin{figure}[H]
\centering
\includegraphics[width=0.6\linewidth]{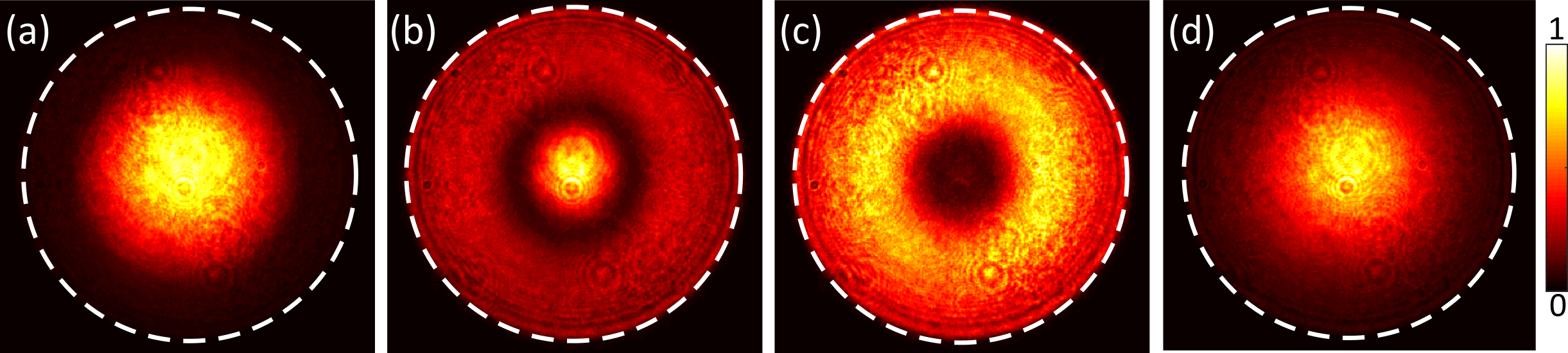}
\caption{Measured Fourier plane images in reflectance from the thin film device with central wavelength set to $598$ nm (a), $632$ nm (b), $640$ nm (c) and $694$ nm (d) with dashed lines indicating $\mathrm{NA}=0.9$.}
\label{fig:otf}
\end{figure}

\begin{figure}[H]
\centering
\includegraphics[width=0.6\linewidth]{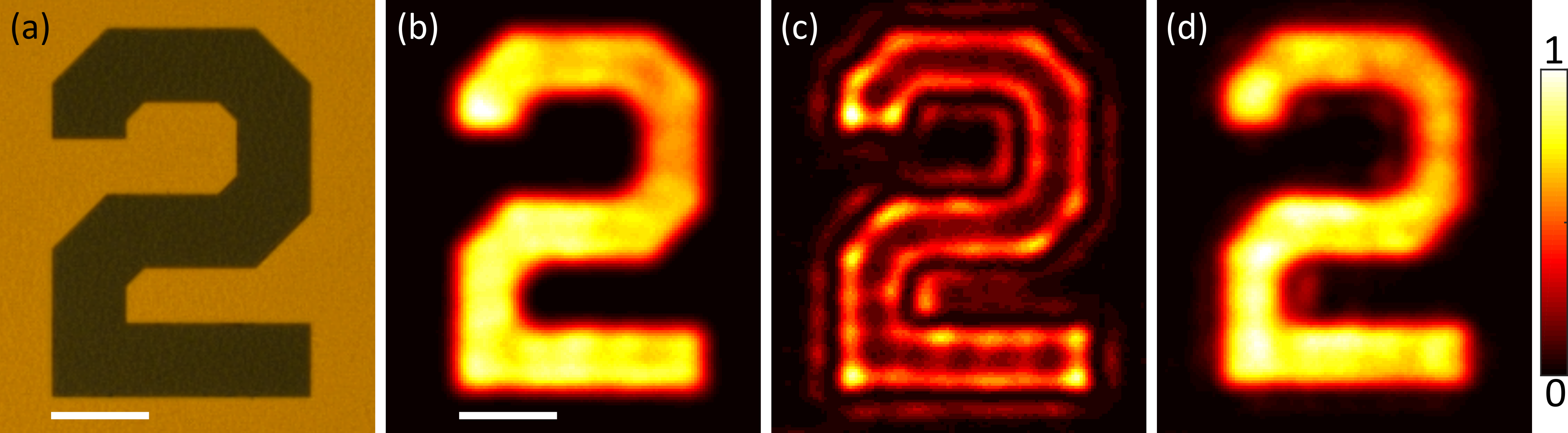}
\caption{Demonstration of edge enhancement in amplitude images reflected from the thin-film mirror.
(a) brightfield microscope image in reflection of experimental amplitude object, experimental reflectance from the thin film mirror at (b) $598$nm, (c) $640$nm  and (d) $694$nm. Scalebar is $100 \mu$m (a), $2 \mu$m (b-d).}
\label{fig:edgedetection}
\end{figure}

\begin{figure}[h]
\centering
\includegraphics[width=0.6\linewidth]{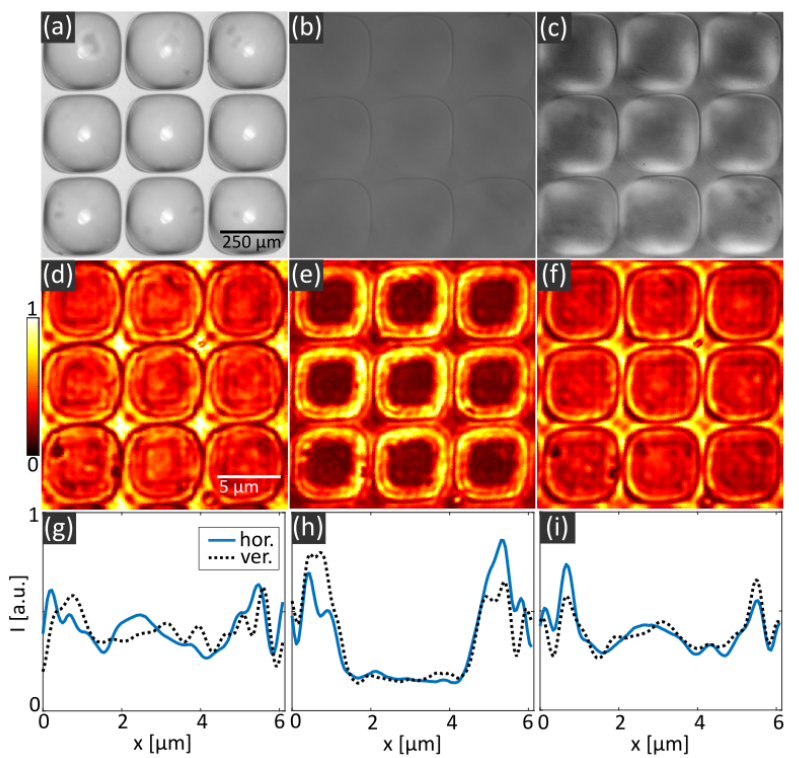}
\caption{Experimental demonstration of phase-gradient detection using a microlens array as a test object. Brightfield image of the object without (a) and with (b) index matching oil, differential interference contrast image of lenses immersed in index matching oil (c). Images reflected from mirror and corresponding line profile plots through the horizontal and vertical center of the central lens at $598$nm (d) and (g), $640$nm (e) and (h) and $694$nm (f) and (i).}
\label{fig:microlens}
\end{figure}

The device was fabricated on a [$100$] p-type silicon wafer. A titanium adhesion layer with a thickness of $3$ nm was deposited on the wafer at $0.2$ \AA/s followed by a $150$ nm thick layer of Au deposited at $0.7$ \AA/s using an IntlVac NanoChrome II e-beam evaporator. A $155$ nm thick layer of SiO$_{2}$ was grown on the sample using an Oxford Instruments PLASMALAB $100$ PECVD system. The sample was then loaded back into the e-beam evaporator and a $31$ nm thick Au film deposited on the SiO$_{2}$ layer. The reflectance was measured for both \textit{p-} and \textit{s-}polarization as a function of angle of incidence using a spectrometer (Agilent Cary 7000 UV-Vis-NIR Universal Measuring Spectrometer (UMS)) with the results shown in Figure \ref{fig:reflectance}. The design of the spectrometer limits the minimum and maximum measurable angle of incidence to $6^\circ$ and $84^\circ$ respectively.
The measurements show absorption of about $98.5\%$ of the incident intensity around $640$ nm at $6^\circ$ angle of incidence shifting towards the blue as the angle of incidence increases for both \textit{p-} and \textit{s-} polarizations in line with the numerical simulations. The absorption maximum is shifted by approximately $10$ nm compared to the numerical result which we attribute to variations in the dielectric function of gold and silicon dioxide in thin films compared to bulk material as well as an uncertainty of $\pm 10\%$ in deposited film thickness for both fabrication methods.
To experimentally confirm the image processing capability of the thin film device, we demonstrate edge enhancement of an amplitude image and visualize phase gradients in an image with negligable amplitude modulation. A real image of the object is projected onto the surface of the optical Salisbury screen using the benchtop K\"ohler-illumination configuration in Figure \ref{fig:setup}. 
Unpolarized light from a supercontinuum laser source (Fianium SC-450-2) is passed through a fiber-coupled tunable filter (Fianium Superchrome VIS-FDS-MM) set to its minimum bandwidth of $5$nm. The light is guided to the setup via a single mode fiber (Thorlabs SM600) and collimated using a NA 0.3 Nikon LU Plan Fluor objective (MO1). The collimated beam is transmitted through the object under investigation which is then imaged onto the sample surface via a telescope consisting of a \textit{f}$=100$ mm lens (Thorlabs LA1509-A) (L1), a beamsplitter (Thorlabs CM1-BS013) and an infinity corrected NA 0.90 Nikon Plan NCG x100 objective (MO2) with a working distance of $1$ mm. 
This combination effectively enables 50x demagnification of the image thus broadening its spatial frequency distribution in Fourier space. The light reflected from the device is imaged onto the camera (L2). As a proof of concept Fourier plane images were obtained using this setup without object and lens (L1) removed as shown in Figure \ref{fig:otf}. These results are consistent with the angular measurements obtained using the spectrometer shown in Figure \ref{fig:reflectance}. The suppression of low spatial frequencies at $640$ nm is apparent. It should be noted that the structure filters higher spatial frequencies at shorter wavelengths (Fig. \ref{fig:otf}(b)). It is therefore possible to tune the optical transfer function of the device by changing the wavelength.
To demonstrate edge enhancement an amplitude image was formed using the figure '2' of a negative USAF-$1951$ resolution test target (Thorlabs R1DS1N) (Fig. \ref{fig:edgedetection}(a)). The results indicate enhanced edges at the absorption wavelength of the absorber (Fig. \ref{fig:edgedetection}(c)) not present at shorter (Fig. \ref{fig:edgedetection}(b)) or longer wavelengths (Fig. \ref{fig:edgedetection}(d)). 
The experimental results are consistent with the numerical findings. It should be noted that the quantitative edge enhancement contrast is sensitive to the bandwidth of the illumination as well as the spatial frequencies generated by the object and those captured by the imaging system. These depend on the sharpness of the edges of the resolution test pattern and the projected image, which are experimentally difficult to quantify.
The spectral bandwidth of the illumination implies an effective optical transfer function that is averaged over the bandwidth of the laser system thereby reducing the contrast between reflected high- and low spatial frequencies.
To confirm the potential for the Salisbury screen to visualize phase gradients, we used an array of PMMA microlenses (FINLENS Flyeye 80 LPI) immersed in index matching oil (Cargille, $n_{\mathrm{oil}}=1.518$). Figures \ref{fig:microlens}(a) and (b) show bright field images of the microlens array with and without the use of matching oil. While the curvature of the lenses immersed in matching oil becomes almost invisible in the brightfield image, it is still apparent in a differential interference contrast (DIC) image of the same sample in Figure \ref{fig:microlens}(c). 
Figure \ref{fig:microlens} (e) shows reflected images of the microlens array at the absorption wavelength of the Salisbury screen. While the central part of the lenses, associated with low phase gradients, remains dark, the outer parts appear bright indicating a steep phase gradient in line with the DIC image in Figure \ref{fig:microlens} (c). At shorter (Fig.\ref{fig:microlens} (d)) and longer wavelengths (Fig.\ref{fig:microlens} (f)) the lens regions appear almost uniform apart from bright lines due to residual scattering at the edges. Horizontal and vertical line profile plots (Fig. \ref{fig:microlens}(g-i)) through the center of the central lens further highlight the intensity contrast generated from the phase gradient. The line profiles have been averaged over an interval of $\pm0.28\mu$m around the horizontal (vertical) center of the lens in order to reduce noise. The results confirm experimentally the ability of the device to convert a phase gradient into a readily measured intensity modulation in the reflected image permitting, in principle, phase imaging of biological cells.
\section{Conclusion}
In conclusion we have demonstrated both computationally and experimentally the all-optical, real-time image processing capability of a two layer thin film absorber structures using the example of a Au/SiO$_2$/Au multilayer Salisbury screen configuration. We have characterized the wavelength-selective angular response of the device in reflectance and demonstrated its highpass- spatial frequency filtering potential. We have utilized this  to demonstrate edge enhancement of an amplitude image reflected from the surface of the device and demonstrated its ability to convert phase gradients into intensity variations in the reflected image. The structure presented here supports spatial frequency filtering at a wavelength in the visible spectral range tunable by the choice of the geometric parameters offering the intriguing possibility of being further able to design highly specialized devices for application-dependent integrated optics. More complex layer structures with additional absorption lines could for example have potential for processing color images. 
Applications of the presented research lie in the field of live-cell imaging and biological and material imaging approaches where the thin film absorber could provide cheap and ultra compact alternatives to bulky phase imaging approaches.

\section*{Funding Information}
Australian Research Council Discovery Projects (DP160100983) and Future Fellowship scheme (FT140100514).
\section*{Acknowledgments}
This work was performed in part at the Melbourne Centre for Nanofabrication (MCN) in the Victorian Node of the Australian National Fabrication Facility (ANFF).

\footnotesize{\bibliography{readcube_export,Additional_literature,phdetec_library}}
\bibliographystyle{ieeetr}
\end{document}